\documentclass[aps,prl,twocolumn,lengthcheck]{revtex4}
\usepackage{graphicx}

\begin{document}


\title{The $^{15}$O($\alpha$,$\gamma$)$^{19}$Ne Breakout Reaction and Impact on X-Ray Bursts}

\author{W. P. Tan}
\email{wtan@nd.edu}
\author{J. L. Fisker}
\author{J. G{\"o}rres}
\author{M. Couder}
\author{M. Wiescher}
\affiliation{Department of Physics, University of Notre Dame, Notre Dame 46556 }


\begin{abstract}
The breakout reaction $^{15}$O($\alpha,\gamma$)$^{19}$Ne, which regulates
the flow between the hot CNO cycle and the rp-process, is critical for the explanation of the burst amplitude
and periodicity of X-ray bursters. We report on the
first successful measurement of the critical $\alpha$-decay branching ratios of relevant states in $^{19}$Ne
populated via $^{19}$F($^3$He,t)$^{19}$Ne. Based on the experimental results and our previous lifetime measurements of these states, we derive the first experimental rate of  $^{15}$O($\alpha,\gamma$)$^{19}$Ne.
The impact of our experimental results on the burst pattern and
periodicity for a range of accretion rates is analyzed.
\end{abstract} 

\maketitle

An X-ray burster is characterized by a repeated sudden
increase of X-ray emission within only a few seconds to a total
energy of about 10$^{39\textrm{--}40}$ ergs~\cite{Woosley04}. The recurrence time between single
bursts can range from hours to days. The characteristics of X-ray burst phenomena are
being studied extensively today using a number of space-based
X-ray observatories such as RXTE, BeppoSAX, Chandra, HETE-2, and
XMM/Newton. More than eighty galactic sources of X-ray bursts have
been identified since their initial discovery in 1976~\cite{Galloway06}.
These bursts are explained as thermonuclear explosions in the atmosphere of an accreting
neutron star in a close binary system~\cite{Woosley76,Joss77}.
When critical values for density and temperature are reached in
the neutron star atmosphere, the freshly accreted hydrogen and
helium ignites and burns via the hot, $\beta$-limited CNO cycles at a constant rate~\cite{Wiescher1999}. Depending on the strength of the
$^{15}$O($\alpha,\gamma$)$^{19}$Ne reaction, break-out from the
hot CNO cycles will occur, fueling the rapid proton capture
(rp)-process~\cite{Wallace1981,Schatz1998}. The rp-process converts the light element
fuel into heavy elements from Fe-Ni up to Cd-Sn
within only a few seconds.

The strength of the $^{15}$O($\alpha,\gamma$)$^{19}$Ne break-out
reaction not only regulates the ignition point of the actual burst
but affects also the burst recurrence rate. However,
the large experimental uncertainties of the
$^{15}$O($\alpha,\gamma$)$^{19}$Ne reaction rate prevented the use
of the rate as a tool to identify the conditions required for the
ignition and the recurrence time of the bursts. Quite the
opposite, a comparison between astronomical observations and
theoretical model predictions was used to constrain the reaction
rate~\cite{Cooper06,Fisker06}. Fisker et al. demonstrated that the
previous speculative estimate on the lower limit of the reaction rate would lead to
burst quenching for observed accreting rates of $\dot{M}\sim
10^{17} \textrm{g}\,\textrm{s}^{-1}$~\cite{Fisker06}. This
suggested that the previous lower limit estimate was too low.
Cooper et al. on the other hand performed an analytical stability
study suggesting that the widely used rate of Langanke et al.~\cite{Langanke1986} might be too high~\cite{Cooper06}. An
experimental study of $^{15}$O($\alpha,\gamma$)$^{19}$Ne is
clearly necessary. It will not only reduce the wide range of
uncertainty on the actual ignition conditions but may also provide
new constraints on accretion itself by identifying the transition
point between the thermonuclear runaway in the burst~\cite{Fisker03} and steady state burning as expected for higher
accretion rates~\cite{Schatz99}.

There have been many attempts in the past to provide experimental
data for determining the $^{15}$O($\alpha,\gamma$)$^{19}$Ne
reaction rate. While progress has been made, the presently used
rates are still relying mainly on theoretical estimates~\cite{Langanke1986,Wiescher1999}. The $^{15}$O($\alpha,\gamma$)$^{19}$Ne reaction
rate is expected to be dominated by a single resonance level at
an excitation energy of 4.03 MeV in $^{19}$Ne~\cite{Langanke1986} where the resonant rate is defined by $N_{A}<\sigma v>_{res}\propto(kT)^{-3/2}(2J+1)\Gamma_{\gamma}\Gamma_{\alpha}/\Gamma\exp(-E/kT)$. 
Present intensities of radioactive $^{15}$O beams are insufficient for a direct measurement.
Therefore, past studies have focused mainly on the use of indirect techniques by measuring the
characteristic nuclear structure features of the $^{19}$Ne
compound nucleus for the determination of the resonance parameters for
the reaction rate~\cite{Magnus1990,Mao1995,Hackman2000,Laird2002,Davids2003a,Rehm2003,Visser2004,Tan2005}.
Yet none of the studies were successful in determining a model
independent reaction rate with well defined experimental limits.

Recently excitation energies and $\gamma$ partial widths $\Gamma_{\gamma}$ or the inverse of lifetimes of the states in $^{19}$Ne near $\alpha$-threshold have been measured with the improved Doppler shift attenuation method~\cite{Tan2005}. The result on the lifetime of the 4.03-MeV state was also confirmed in an independent measurement at TRIUMF~\cite{Kanungo2006}. The
remaining quantity to be determined is the $\alpha$-decay
branching ratios B$_{\alpha}=\Gamma_{\alpha}/\Gamma$ that have
mostly been estimated from the $\alpha$ strengths of the mirror
states in $^{19}$F with large systematic model
dependent uncertainties inherent to the DWBA analysis of
$\alpha$-transfer reactions on $^{15}$N~\cite{Langanke1986,Mao1995,Oliveira1996}. Several
attempts have been made in the past to measure the relative
$\alpha$-decay widths directly~\cite{Magnus1990,Kurokawa1998,Davids2003a,Rehm2003}. While
this approach was successful for higher-lying states in
$^{19}$Ne, it failed for the critical levels near the $\alpha$
threshold due to the low decay branching.

Here we report, for the first time, the successful laboratory
measurement of $\alpha$ decay of the unbound states in $^{19}$Ne
which provides an experimental rate for
$^{15}$O($\alpha$,$\gamma$)$^{19}$Ne and discuss the astrophysical implications.

\begin{figure}
\includegraphics[scale=1.0]{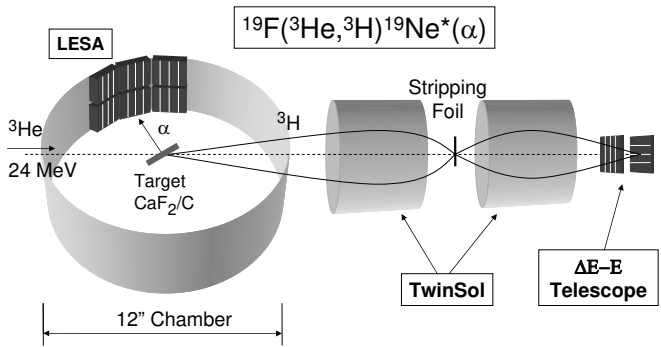}
\caption{\label{fig:setup}Schematic setup of this experiment is shown (not to scale for better presentation).}
\end{figure}

The $^{19}$F($^3$He,t) reaction was used to populate $\alpha$
unbound states in $^{19}$Ne; their $\alpha$-decay branchings were
determined through t-$\alpha$ coincidence measurements. The
detection system was optimized for the measurement of low energy
$\alpha$ particles ($\le$ 1 MeV) and an overall detection
efficiency sufficient for probing branching ratios as low as
$10^{-4}$.

A schematic drawing of the experimental setup is shown in
Fig.~\ref{fig:setup}. The $^3$He beam of 24 MeV was produced at
the FN tandem accelerator of the University of Notre Dame to
bombard a 40 $\mu$m/cm$^2$ thick CaF$_2$ target evaporated on a
20 $\mu$m/cm$^2$ thick Carbon foil. The TwinSol facility at Notre Dame,
a dual in-line superconducting solenoid ion-optical system~\cite{twinsol}, was used as a
large-acceptance momentum separator to separate tritons from other
reaction products. The acceptance range corresponded to a solid
angle of 50 msr. A large area position-sensitive $\Delta$E-E
telescope consisting of two  500 $\mu$m thick silicon detectors
was positioned close to the focal plane of TwinSol to identify and
track tritons.

A Low Energy Silicon-strip Array (LESA) was designed to detect the
low energy $\alpha$ particles from the decay of the excited states
in $^{19}$Ne. The array consists of six identical 300 $\mu$m thick
silicon-pad detectors, each of which has 4 strips and an area of
4x4 cm$^2$. To reduce the detection threshold the dead layer was
limited to a thickness of $<0.05$ $\mu$m. This translates into an
energy loss of $<14$ keV for 200 keV $\alpha$ particles. The particle
identification in LESA was achieved by measuring the particle's energy and time of flight
from the target to LESA.

\begin{figure}
\includegraphics[scale=1.0]{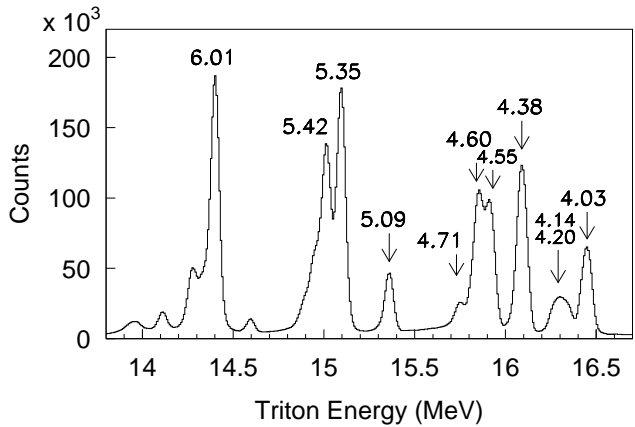}
\caption{\label{fig:de_e}The total energy spectrum of tritons detected in the telescope is shown using the 
$\Delta$E-E particle identification technique. Peaks related to the excited states in $^{19}$Ne are labelled.}
\end{figure}
In Fig.~\ref{fig:de_e}, the triton energy spectrum is presented and
demonstrates the good separation that was achieved for the excited states in $^{19}$Ne. In particular, the 4.03-MeV state is well separated although the two high spin states at 4.14 and 4.20 MeV are not resolved. Furthermore, enough counts were collected to ensure that the sensitivity of the measurements of $\alpha$ decay branching ratios reaches as low as $10^{-4}$. However,
double hits in the telescope inevitably contribute to the background in the triton spectrum in Fig.~\ref{fig:de_e}. In rare cases, double-hit events with high energy electrons can contaminate a given triton peak with tritons from higher-lying states ($>$0.5 MeV higher in energy).
The effect on small $\alpha$ decay branching fractions is demonstrated below.

\begin{figure}
\includegraphics[scale=1.0]{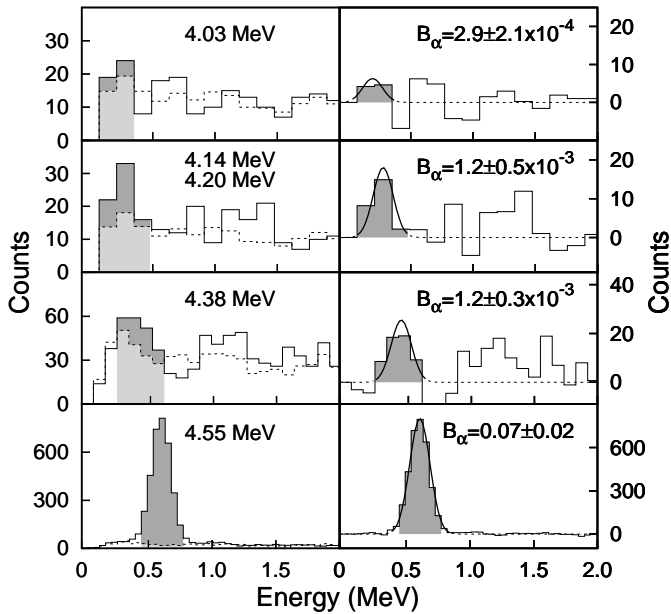}
\caption{\label{fig:levels}Kinematically corrected $\alpha$ energy spectra are shown in coincidence
with the tritons for the lower lying states at 4.03-4.55 MeV. Left panels: the solid histograms indicate all
coincidence events while the dashed histograms show the background normalized from large ensemble of random
coincidence events; the shaded areas present the expected $\alpha$ energy range for the corresponding states
in $^{19}$Ne. Right panels: the net $\alpha$
spectra are shown after the background deduction; the smooth curves are simulated calculations assuming a decay
branching as measured. 
See text for details.}
\end{figure}

A detailed discussion and analysis of the experimental results
will be presented in a forthcoming paper~\cite{Tan2007}. Here we only summarize
our results and focus on the aspects relevant for the
determination of the reaction rate. Kinematically corrected $\alpha$ spectra for the decay of the observed  $^{19}$Ne states are shown in Fig.~\ref{fig:levels} after appropriate triton and timing gates are applied. The total and background $\alpha$ spectra are presented in the left panels while the net spectra are plotted in the right panels. The background was obtained from random coincidence events outside of the $\alpha$ timing gates which have a much larger count rate than those within the gates for rare $\alpha$ decay cases. For this reason, the additional statistical error from the determination of the background does not contribute significantly to the uncertainty of the net spectra. The background rise to the low energy end is due to the energy loss of high energy electrons in LESA.
The coincident $\alpha$s from the corresponding states are indicated by the shaded areas while 
the structures at higher energy are $\alpha$s leaking from higher-lying states with much larger B$_{\alpha}$s because of double hit events with high energy electrons. Fortunately, the double-hitting electrons deposit at least 0.5 MeV in the telescope and the resulting 0.5 MeV gap is sufficient to separate these $\alpha$s from
the weak $\alpha$ decay signals for the states at 4.03-4.38 MeV.

While the $\alpha$ group for the decay of the 4.55-MeV state in
$^{19}$Ne is quite pronounced -- similar to the decay of the higher
excited states at 4.60, 4.71, and 5.09 MeV -- only weak
$\alpha$-decay branching ratios on the order of $10^{-3}$ or less
are observed for the levels near the $\alpha$ threshold.

The present experiment yields for the first time branching ratios
for the states at 4.03-4.38 MeV near the alpha threshold. An
$\alpha$-decay branching ratio of $2.9\pm2.1\times10^{-4}$ was
measured for the 4.03-MeV state consistent with previous
upper limits of $<4.3\times10^{-4}$~\cite{Davids2003a} and $<6\times10^{-4}$~\cite{Rehm2003}. The two states at 4.14
and 4.20 MeV could not be resolved, but a combined branching
ratio of $1.2\pm0.5\times10^{-3}$ was determined. This is
surprisingly large compared to previous predictions and assessments~\cite{Langanke1986,Rehm2003}. 
The measured $\alpha$ peak seems to be lower in energy than the simulated one indicating that these decay events are more likely from the 4.14-MeV state.
As for the 4.38-MeV state, the
present result of $1.2\pm0.3\times10^{-3}$ agrees with the stringent upper limit of  $<3.9\times10^{-3}$~\cite{Davids2003a} while it differs from previously given values of  $0.044\pm0.032$~\cite{Magnus1990} and  $16\pm5\times10^{-3}$~\cite{Rehm2003} which were handicapped by poor
statistics and lack of experimental resolutions. For the states at
higher excitation energies of 4.55-5.09 MeV our new measurements
show excellent agreement with the previous results~\cite{Magnus1990,Laird2002,Davids2003a,Rehm2003,Visser2004}.

\begin{figure}
\center
\includegraphics[scale=1.0]{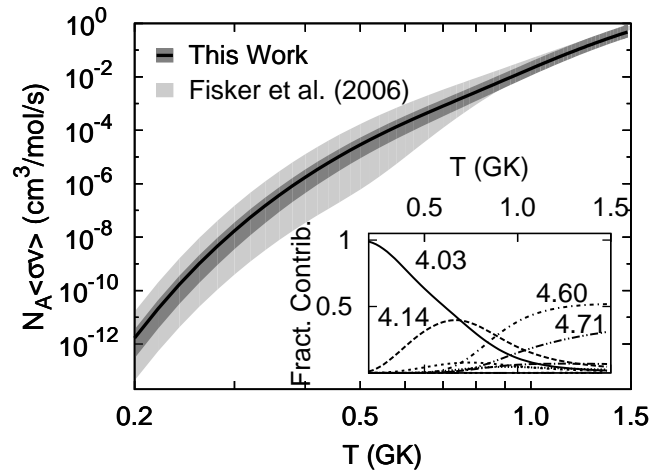}
\caption{\label{fig:rate} The new $^{15}$O($\alpha$,$\gamma$)$^{19}$Ne
reaction rate (black line) is plotted with one sigma uncertainty
indicated by the dark-grey area. The uncertainty range discussed by
Fisker et al.~\cite{Fisker06} are indicated by the light-grey area which
did not take into account the uncertainty from high-lying ($>$ 4.4
MeV) states that contribute significantly at T$>$0.6 GK. The
inset displays the fractional contributions of the individual
resonance states to the reaction rate.}
\end{figure}

The reaction rate for
$^{15}$O($\alpha$,$\gamma$)$^{19}$Ne was directly calculated from the
measured $\alpha$-decay branching ratios and the lifetimes~\cite{Tan2005} of the $\alpha$-unbound states in $^{19}$Ne.
Fig.~\ref{fig:rate} shows the reaction rate as a function of
temperature. The upper and lower limits of the rate are based on
the experimental uncertainties of the branching ratio and lifetime
measurements. Also shown for comparison is the range of
previous theoretical uncertainty~\cite{Fisker06}.

The new experimental rate allows not only a better identification of the ignition
conditions of X-ray bursts but permits also the improved analysis
of the dynamics and mechanism of X-ray bursts. In this context the
accretion rate corresponding to the transition point between
steady state and unstable burning is of particular interest.

The impact of the $^{15}$O($\alpha$,$\gamma$)$^{19}$Ne rate has been
investigated in the framework of a dynamical and self-consistent
spherically symmetric X-ray burst model~\cite{Fisker06}. This model
couples a modified version of a general relativistic hydrodynamics
code~\cite{Liebendoerfer02} with a generic nuclear reaction
network~\cite{Hix99} using the operator-split method. By solving the general relativistic equations, the nuclear reaction flow, the conductive, radiative, and convective
heat transport was computed in a spherically symmetric geometry.
More detailed discussions of the model can be found in a forthcoming paper by Fisker et al~\cite{Fisker2007}.

The transition between unstable burning and steady state burning
is easily observable by the change in burst pattern. The
associated accretion rate, however, correlates only on average
with the observed accretion luminosity~\cite{Lewin87}. The
influence of the nuclear trigger processes on the actual
transition point can therefore only be investigated in the
framework of a theoretical calculation.

\begin{figure}
\center
\includegraphics[scale=1.0]{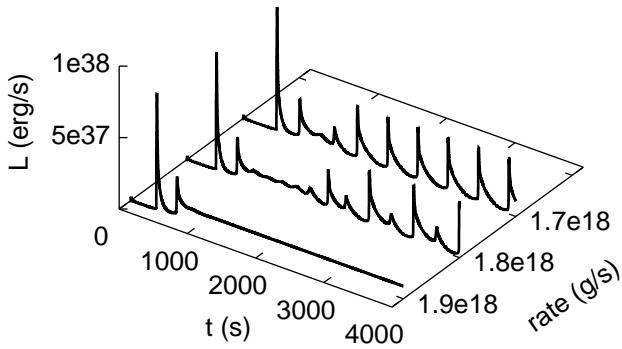}
\caption{\label{fig:wrec_Lt}We used the newly measured rate to calculate the
luminosity originating from the nuclear burning as a function of
time for different accretion rates. As is seen from the constant
luminosity on the graph, the burning is stable for
$\dot{M}\geq 1.9\times 10^{18} \textrm{g}\,\textrm{s}^{-1}$.
}
\end{figure}

Previous simulations~\cite{Fisker03,Heger06} and calculations~\cite{Fushiki87} have determined the transition point to be around
$\dot{M}\sim 2.1\times 10^{18} \textrm{g}\,\textrm{s}^{-1}$ for a
fiducial neutron star with $R=10$ km and $M=1.4M_\odot$ adopting a
solar composition for the accreted matter. However, the importance
of the uncertainty of the $^{15}$O($\alpha,\gamma$)$^{19}$Ne
reaction was not taken into account. With the new measurement, we
are now able to determine the transition point between
steady state burning and unstable burning with significantly
improved accuracy. We performed several calculations for different
accretion rates while tracking the luminosity resulting from the
nuclear burning. The results shown in Fig.~\ref{fig:wrec_Lt}
depict how the burning becomes stable for $\dot{M}\geq 1.9\times
10^{18} \textrm{g}\,\textrm{s}^{-1}$.

We performed identical calculations for the one sigma upper and
lower limits as shown in Fig.~\ref{fig:rate}. The upper limit
yields the same transition accretion rate, whereas the lower limit
increases the transition point to $\dot{M}\approx 2.1\times
10^{18} \textrm{g}\,\textrm{s}^{-1}$. The
uncertainty in the determination of the accretion rate at transition point is thus
about 10\% compared to previous uncertainties of one
order of magnitude~\cite{Fisker06}. Further model studies of the
transition accretion rate are necessary to take better into
account the mass and radius of the neutron star as well as the
accreted composition.

This work shows the importance of laboratory
results for providing stringent limits for the burning conditions
in stellar objects. It demonstrates how experimental nuclear data
can complement observational results and provide important insights
for the astrophysical model simulations. The $^{15}$O($\alpha,\gamma$)$^{19}$Ne
reaction is the key for our understanding of the onset of X-ray
bursts. The experimental results bring us closer to a better
understanding of the complex interplay between fuel supply and
burning processes at the extreme conditions of the neutron star
atmosphere.

\begin{acknowledgments}
We thank M. Beard, A. Couture, S. Falahat, L. Lamm, P. J. LeBlanc, H.Y. Lee, S. O'Brien, A. Palumbo, E. Stech and E. Strandberg for help during the course of the experiment.
This work is supported by the National Science Foundation under
grant No. PHY01-40324 and the Joint Institute for Nuclear Astrophysics (www.jinaweb.org), NSF-PFC under grant No. PHY02-16783.
\end{acknowledgments}

\end{document}